\documentclass[conference]{IEEEtran}
\IEEEoverridecommandlockouts
\usepackage{cite}
\usepackage{amsmath,amssymb,amsfonts}
\usepackage{algorithmic}
\usepackage{graphicx}
\usepackage{textcomp}
\usepackage{xcolor}
\usepackage{longtable}
\usepackage{pdflscape}
\usepackage{url} 
\def\BibTeX{{\rm B\kern-.05em{\sc i\kern-.025em b}\kern-.08em
    T\kern-.1667em\lower.7ex\hbox{E}\kern-.125emX}}
\begin{document}

\title{Integrated Control of Robotic Arm through EMG and Speech: Decision-Driven Multimodal Data Fusion\\
}

\author{\IEEEauthorblockN{Tauheed Khan Mohd}
\IEEEauthorblockA{\textit{School of Information Security and Applied Computing} \\
\textit{Eastern Michigan University}\\
Ypsialnti, Michigan, USA \\
tkhanmoh@emich.edu}

\and
\IEEEauthorblockN{Ahmad Y Javaid}
\IEEEauthorblockA{\textit{EECS Department} \\
\textit{The University of Toledo}\\
Toledo, Ohio, USA \\
ahmad.javaid@utoledo.edu}

}

\maketitle

\begin{abstract}
Interactions with electronic devices are changing in our daily lives. The day-to-day development brings curiosity to recent technology and challenges its use. The gadgets are becoming cumbersome, and their usage frustrates a segment of society. In specific scenarios, the user cannot use the modalities because of the challenges that bring in, e.g., the usage of touch screen devices by elderly people. The idea of multimodality provides an ease to accessing devices of our daily use through various modalities. In this paper, we suggest a solution that allows the operation of a microcontroller-based device using voice and speech. The model implemented will learn from the user's behavior and decide based on prior knowledge. 
\end{abstract}


\section{Introduction}
With the advancement of technology, daily lives become more occupied with devices; the recent development allows users to access electronic devices using several methods and modalities. Cellphones, IoT devices, and smartwatches are all operated by either touch or human speech. Gestures do not play a crucial role in controlling devices in the current scenario. Moreover, if the feature of gestures is exploited, then we can control, operate, and manage various electronic devices throughout the day. From switching off the lights to closing the garage door, all these activities could be controlled by the limb's movement. The idea we have planned to work on is to incorporate multiple modalities while using the technology for daily use. The software developed is a prototype of heavy robots used in industry for several jobs and accomplishing tasks in less time.
Moreover, the same concept can easily be applied to our daily activities using the microcontroller. Multimodal Human-Computer Interaction (MMHCI) is used to decide how we can make computer technology more usable by people, which perpetually requires the understanding of at least three things: the user who interacts with it, the system (the computer and its usability), and the communication between the user and the system \cite{jaimes2005multimodal}. In our research, we have surveyed the usage of Multimodality in the myriad industry and daily use applications. For the extension of our work, we decided to use electromyography as the input for the application. To our knowledge, no application has used electromyography with human speech to generate commands for the robotic arm. Moreover, we have applied several machine learning techniques to determine which works best with our application.

\section{Related Work}

The usage of technology is managed by several modalities available to access them. IoT-based devices are controlled by phone through touch or speech commands. Decision-level fusion involves the fusion of sensor information that is preliminarily determined by the sensors. Examples of decision-level fusion methods include weighted decision methods, classical inference, Bayesian inference, and Dempster–Shafer method \cite{llinas1998introduction}. The fuzzy clustering algorithms, which involve assigning data points to clusters, while items belonging to different clusters are as dissimilar as possible, are proposed to authenticate a person \cite{chatzis1999multimodal}. The decision-based fusion for heterogeneous sensors for real-time monitoring is implemented using Bayesian clustering \cite{beyca2015heterogeneous}. Similar techniques are used in the pan-sharpening of images for remote sensing. Two renowned techniques are used for fusion remote images, viz. À trous wavelet transform-based pan sharpening (AWLP) and À trous wavelet transform followed by context-based decision fusion (AWCBD). The latter one is good for sharpening of larger objects in the image in suburban areas. In the 2006 IEEE Fusion contest, the Institute of Electrical and Electronics Engineers, À trous wavelet fusion (AWLP) and the Laplacian pyramid-based context decision fusion (CBD) were rated as the winner as they both had better results in urban areas \cite{luo2012decision}. 

The fusion of three-dimensional data from cultural sites was analyzed to achieve comprehensive surveys. The idea is to combine full-resolution laser scanning and photogrammetry. Three-dimensional data streams were generated using Light detection and ranging (LiDAR) \& ground-based images were captured through drones. LiDAR is a method for calculating distances by measuring the time a laser takes to reflect back to a sensor (this technology can be used in mobile devices nowadays) \cite{reutebuch2005light}. The data is fused to generate a holistic view, and the experiment was performed on an ancestral church. The fusion implemented is based on the capturing of two data collection techniques of images LiDAR and structure from motion (SfM), a photogrammetric range imaging technique which is used to create three-dimensional digital structures from two-dimensional series of images \cite{schonberger2016CVPR}. The LiDAR captures the room's interior, while SfM captures the exterior walls. The data fusion algorithm technique is not explicitly explained, and no empirical proof was provided to evaluate the implemented techniques \cite{hess2015fusion}\cite{dalla2015challenges}.  

Remote sensing is one of the crucial areas where multimodality is used extensively. It uses three types of fusion, viz. raw-data level, feature level, and decision level. The decision-level fusion combines results from different modalities, provided they complement each other. One can expect to increase the robustness of the decision through fusion in comparison with the results obtained independently. The achieved fusion of results is confirmed due to consensus. In case of discordance, the result is concluded based on majority voting. Decision-based fusion approach used in pattern recognition \cite{dalla2015challenges}. 

Another approach for predicting the probability of events is through the stochastic model describing a sequence of possible events in which the probability of each event is contingent only on the state achieved in the preceding event, and this method called Markov Chain \cite{blanchet2016markov}. 

The classification of data for fusion techniques can be divided into four broad categories
\begin{enumerate}
\item 	The relation between the input data sources. These relations are defined as (a) complementary, (b) redundant, or (c) cooperative data.
\item 	According to the input/output data types and their nature.
\item 	Following an abstraction level of the employed data, which are (a) raw measurement, (b) signals, and (c) characteristics or decisions.
\item 	Dependent on the architecture type: (a) centralized, (b) decentralized, or (c) distributed \cite{castanedo2013review}. 
\end{enumerate}

The decision-based fusion is classified into two broad categories: the first one is the Bayesian Method, and the second one is the Dempster-Shafer Theory. 
The Bayesian Method, rooted in probability theory, serves as a powerful framework for updating and refining our beliefs in the face of new information. In technical terms, it enables the calculation of posterior probabilities by combining prior beliefs with the likelihood of observed data. This method provides a systematic and mathematically rigorous approach to inference, particularly in predictive modeling and decision-making scenarios. Its flexibility makes it applicable across various domains, from machine learning to statistical analysis, where dynamically adjusting our understanding based on incoming data is paramount.

Whereas the Dempster-Shafer Theory, in the realm of uncertainty and evidence fusion, the Dempster-Shafer Theory (DST) stands out as a robust framework. Developed to handle situations where multiple sources of evidence contribute to a decision-making process, DST utilizes belief functions to represent the degree of confidence or uncertainty associated with each piece of evidence. By combining these belief functions, DST offers a principled way to manage and process diverse, sometimes conflicting, information. This makes it a valuable tool in fields such as artificial intelligence, decision support systems, and forensic analysis, where complex decision-making requires careful consideration of uncertain and potentially conflicting sources of evidence.

\cite{castanedo2013review}.

Multimodal fusion is the heart of any multimodal sentiment analysis engine. There are two main fusion techniques: feature-level fusion and decision-level fusion. Feature-level fusion is implemented by concatenating the feature vectors of all three modalities to form a single long feature vector. Despite its simplicity, this method produces accurate results. We concatenated the feature vector of each modality into a single feature vector stream. This feature vector is used for classifying each video segment into sentiment classes. To estimate the accuracy, we used tenfold cross-validation. 

In another experiment, a quadcopter was controlled by Myo armband. The proposed solution used real-time hand gesture recognition for flight control of multiple quadrotors. It used electromyography signals (EMG) and Convolutional Neural Networks (CNN) to simplify flight operation control and make it more intuitive for the user. These days unmanned aerial vehicles are used in civil and military fields. The Myo armband exploited in the experiment consists of an 8-channel dry-electrode with a low sampling rate (200 Hz) integrated with Bluetooth for communication.
Moreover, the Myo sensor includes an inertial measurement unit (IMU) with 9 degrees of freedom (DoF). The convolutional neural network is implemented using the library implemented in Python called TFlearn. It is a transparent deep-learning library implemented in modules and built on top of TensorFlow. The TFLearn is designed to offer a higher-level API to TensorFlow to facilitate and speed up experiments. The architecture implemented was selected by trial and error based on the slow fusion model \cite{redrovan2018hand}.

MYO Mapper is a free, open-source, cross-platform application that maps EMG data into Open Source Control (OSC) messages. It is developed using MYO open-source API, and it helps musicians explore the possibility of creating musical interfaces using MYO armbands.  The muscle activity in our body is analyzed via EMG data originating from the somatic nervous system and transported to the muscles through various nerves. The feedback is in the form of digital guidelines for effectively using the MYO Mapper to manage EMG data and IMU data from MYO in collaboration with a machine learning tool called Weakinator. This tool allows users to use machine learning for building novel musical instruments, gestural game controllers, computer vision or computer listening systems, and more \cite{di2018myo}.

Hand gestures have various applications in the engineering and medical fields. The foremost problem with the gesture is identifying the input in real-time. In another experiment, a MYO armband commercial sensor labeled the output. The model proposed uses k-NN and a dynamic warping algorithm. To evaluate the performance of the proposed model five classes of gestures are reviewed for accuracy. The result shows after performing the experiments, it can find 86\% accurate gestures than the MYO system, which is 83\% as per their performed experiments.  The experiment comprises five stages: signal acquisition, preprocessing, feature extraction, classification, and post-processing. The experiments are performed several times to train the system in this process. The gesture is recognized based on the trained model, the empirical data is collected, and it performs well compared to conventional MYO gesture recognition software. The data mentioned justifying it is in synchronization with the data portrayed in our experiments for calculating the performance of the MYO armband while using it with speech commands \cite{benalcazar2017hand}.

The gestures are represented by Electromyography data, which is the recording of the electrical activity of the muscle tissue or its representation as a visual display or audible signal, using electrodes attached to the skin or inserted into the muscle. The challenge with electromyography data is that it is hard to club with deep learning algorithms; the primary challenge is to generate the voluminous data. Then, a model should be trained to exploit the user’s data. In another experiment, two datasets of 18 and 17 non-disabled participants are captured using a low-cost, low-sampling rate (200Hz), 8-channel, consumer-grade, dry electrode MYO armband is used. In machine learning, a convolutional neural network is a class of deep, feed-forward artificial neural networks most commonly applied to analyzing visual imagery.  A convolutional neural network (CNN) is augmented using the transfer learning mechanism to utilize the data from the initial dataset used for training. The result shows that the proposed model achieves the same accuracy as a joystick implemented on a robotic arm. The robotic used has a degree of freedom in 6 operations using the servo motors. The CNN achieves an average of 97.81\% accuracy on seven hand gestures from 17 participants from the second dataset datasets. The data processing of CNN for gesture classification shifts the focus from engineering to feature learning. The feature extraction is a part of CNN; the calculations required before feeding the input to the proposed model have been reduced significantly.
Moreover, the need for the proposed algorithm to execute on an embedded system requires substantial efforts to reduce the initial processing effort. The result was reduced to calculating the spectrograms of the raw sEMG data. The spectrograms are used to store non-stationary signals from sEMG data. For each sensor of the MYO armband, 52 samples (260ms) per spectrogram was leveraged using Hann windows of 28 points per FFT and an overlap of 20. The matrix yields 8*4*15 (channels * Time bins * frequency bins). The matrix is then rearranged to make it a suitable feed for the CNN.
The new matrix is 4*8*15. Finally, the baseline drift and motion artifact, the lowest frequency bin, were removed, and the finalized matrix was 4*8*14. The famous architecture for extracting information is called the slow fusion model, where temporal data start on disconnected parallel layers and slowly fuse across the network, which results in more temporal data for the higher layers. A similar model can be implemented successfully for Multiview image fusion. The matrix used for CNN for the fusion of images is represented as (Time * Spatial * Frequency), eventually representing non-stationary information. The foremost hypothesis of this experiment is the extraction of information from rich data of sEMG when subjects are performing gestures. The idea of using transfer learning for CNN training is that pre-training allows the model to capture more robust features. The CNN then can use general features to prepare an effective model of a new subject sEMG data \cite{cote2017transfer}.

Several applications of MYO armband are found over the internet. Their usage varies from gaming control to managing Unmanned Aerial Vehicles (UAV) flights. Hand cricket is exploited to classify the data captured using the MYO armband. The proposed system is trained using the real-time data fed in MATLAB, which determines the probability of winning the game. The algorithm used in this experiment is a Support Vector Machine (SVM), which helps to train the data and achieve the accuracy of 92\% and 84\%, respectively. The data flow starts from the Raw format, a matrix of (K-Samples * M Points * N Channels) sent for pre-processing. The third stage required feature extraction and a model created with training set data; once the training set created; the test set data was applied to judge the model's accuracy. Pre-processing of data includes DC-offset removal and Notch filter. The DC component in the signals doesn’t align with mean zero, creating a problem during the Fourier transformation. Whereas the notch filter is used to filter frequency with the range, here in this experiment, the notch filter is set to 60Hz, and it is used to remove the power line noise from the obtained raw signal \cite{krishnan2017recognition}. The third stage of training the dataset is feature extraction. In this stage, several mathematical operations are applied to each signal received. These operations are simple square integral, maximum, minimum, mean frequency, and mean power. The fourth stage is using a Support Vector Machine (SVM) to classify the data after pre-processing. The accuracy of SVM is judged when the trained data set can classify the gesture accurately. The results examined suggest the trained model can classify gestures 92.4\% for player 1 and 84.27 for the second player with 3.075 and 3.3325 as their standard deviation respectively \cite{krishnan2017recognition}.

In the past year, there have been several technical papers published evaluating the performance of the MYO Armband. The use of MYO armband, as mentioned, is not only applicable to engineering but has various applications in bioengineering. One such experiment was performed in Denmark by bio-engineering students. This experiment aimed to compare the MYO armband’s narrow bandwidth with a conventional EMG acquisition system, which captures the full spectrum to assess its usability for pattern recognition. The MYO armband is a low-cost sensor; it does not have an extensive feature to capture the complete bandwidth like a professional EMG acquisition. A crossover study was performed on eight participants performing nine hand gestures. Six features were extracted from the data and classified using Linear Discriminant Analysis (LDA), which are six Myo band electrodes to cover the following anatomical landmarks 2 cm distal to the elbow: extensor carpi ulnaris, ulna, extensor digitorum, extensor carpi radialis longus, flexor carpi radialis, and palmaris longus. The empirical data achieved explains the classification error of 5.85±3.63\% for the Conventional EMG Acquisition system, while the MYO Armband has an accuracy of 9.86±8.05\% with no significant difference. Despite low bandwidth, it concludes the MYO Armband is suitable for pattern recognition applications. The data is acquired with Mr. Kick software at 2KHz sampling frequency with analog filtered between 10-500Hz. While the MYO armband data using MYO SDK MATLAB Mex Wrapper toolbox via Bluetooth 4.0, \cite{mendez2017evaluation}.

\section{Machine Learning Implementation for Speech and EMG Data}
Designing a system with multiple modalities brings challenges and complexities. The system design with several input methods requires a decision mechanism. There are several approaches available, which include the implementation of machine learning and Artificial Intelligence techniques. Machine learning is divided into two broad categories: supervised learning and unsupervised learning. Supervised learning algorithms learn and decide the system's decision to work efficiently. It learns based on data provided initially and then gradually learns while the system is in use. The other approach is unsupervised learning algorithms based on the categorical classification of massive unlabeled data. The users do not need to supervise the model since the agent works independently.

It is interesting to implement supervised machine learning algorithms and techniques to make the robotic arm capable of learning from its previous decisions. Based on previous decisions and gestures, the system will learn to execute a movement and enhance this throughout the experiment. We need the system to correctly analyze and evaluate the electrical signals of the muscle tissue in real-time. Therefore, we are interested in supervised learning for classification and supervised learning from regression techniques.
Classification is one type of supervised learning model. This model attempts to conclude observed values. One or more input is provided to the classification algorithm, and this one will try to predict the value with one or more outcomes. The system is trained first with inputs, and accordingly, it takes action later. Several algorithms are available for supervised learning based on speed, accuracy, and data size. Kernel SM, Random Forest, and gradient boosting tree are the algorithms available for precision, while Nave Bayes is used when data is too large. Decision trees and logistic regression are used for processing data quickly [18]. In addition, supervised learning is used for regression to predict and classify outputs from 0 or 1 numerically. For accuracy, the algorithms available are Random Forest, Gradient Boosting Tree, and Neural Network. For speed, decision tree and linear regression are used [18]. 

For our experiment, the input data will be the input data for limbic electrical signals gathered by the MYO armband and voice recognition instructions obtained by the API. An array with the input signals obtained from the sensor and voice commands received from the voice recognition software was created. This will be our training data. This data is composed of 900 interactions. An interaction is defined as either a gesture or a voice command. In addition, another set of data was generated. This one contains the fusion data generated from the training data. The fusion data will be our testing data. Therefore, the model will need to accurately classify, interpret, and label the signals to enhance the fusion data, which will eventually get to the robotic arm. This outcome produced by the agent will be a gesture that the robotic arm will execute. There are only a fixed number of movements that we can detect.

According to these criteria, these are the machine learning techniques models that are available and can be implemented in Python:

\begin{enumerate}
\item Linear Discriminant Analysis: Linear discriminant analysis (LDA) is commonly used as a dimensionality reduction technique in the pre-processing step for pattern classification and machine learning applications. The goal is to project a dataset onto a lower-dimensional space with good class-separability to avoid overfitting (curse of dimensionality) and reduce computational costs [20]. This algorithm is critical for the project since it will maximize the separability of the different categories, maximizing the difference between the means of each of them. At the same time, the variation of each of the categories is minimized so the same categories are grouped together. By doing this, the model will be able to identify each of the data types easily. We have five singular types of EMG signals, and they need to be classified into movements (fist, wave left, wave right, fingers spread, and double tap).  \cite{xanthopoulos2013linear}. 

\item K nearest neighbors algorithm (KNN) is a non-parametric, indolent learning algorithm. Its motivation is to utilize a database in which the information focuses are isolated into a few classes to anticipate the characterization of another sample point, as shown in Figure \ref{fig:KNN_Classifier.png}. The k nearest neighbors have been used widely in pattern recognition. The algorithm is easy to understand conceptually, and the tendency toward error is bounded twice by the Bayes error. The accuracy of K-neighbor surpasses those of sophisticated classifiers. The random subspace method relies on a stochastic process that randomly selects components [21]. We will use the data obtained from the sensor and the voice recognition software as our training data. Our testing data will be the data obtained from the data fusion process. \cite{ho1998nearest}. 

\begin{figure}[h!]
            \centering
            \includegraphics[width=0.4\textwidth]{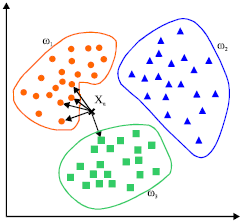}
            \caption{K-NN classifier: Classifying using Knearest neighbors algorithm.png}
            \label{fig:K_NN_Classifer}
\end{figure}

\item Decision Tree Classifier: Decision trees are a type of supervised machine learning (that is, they explain the input and corresponding output in the training data) where the data is continuously split according to a specific parameter. The tree can be explained by two entities: decision nodes and leaves. The leaves are the decisions or the outcomes, and the decision nodes are where the data is split. We will try to predict which of the six gestures will be executed. Again, the data obtained from the EMG sensor and the voice recognition software will be fed into the algorithm. This will be training the data. The data obtained after the fusion process will be the testing data. The agent will filter the data for us and output the actual gesture to be executed.

\item Gaussian Naive Bayes: Gaussian Naive Bayes is a classification algorithm for binary and multi-class classification problems. The technique is easier to understand when described using binary or categorical input values. The approach is called naive Bayes or idiot Bayes because the calculation of the probabilities for each hypothesis is simplified to make their calculation tractable. Rather than attempting to calculate the values of each attribute value P(d1, d2, d3—h), they are assumed to be conditionally independent given the target value and calculated as P(d1—h) * P(d2—H), and so on. This is an extreme assumption that is most unlikely in real data, that is, that the attributes do not interact. Nevertheless, the approach performs surprisingly well on data where this assumption does not hold \cite{john1995estimating}. 

\item Support Vector Machine: Support vector machine (SVM) is a supervised machine learning algorithm used for both classification and regression challenges. However, this is mostly used in classification problems. In this algorithm, we plot each data item as a point in n-dimensional space (where n is the number of features present), with each feature being the value of a coordinate. Then, we perform classification by finding the hyperplane that differentiates the two classes \cite{cortes1995support}. 

\item Logistic Regression:  Logistic regression is a machine learning technique extracted from the field of statistics. It is the go-to method for binary classification problems (problems with two class values). Logistic regression is named for the function used at the core of the method, the logistic function, as shown in Figure \ref{fig:Logistic_Regression}. The logistic function, also called the sigmoid function, was developed by statisticians to describe properties of population growth in ecology, rising quickly and maxing out at the carrying capacity of the environment. It is an S-shaped curve that can take any real-valued number and map it into a value between 0 and 1, but never exactly at those limits \cite{king2001logistic}.

\begin{figure}[h!]
            \centering
            \includegraphics[width=0.5\textwidth]{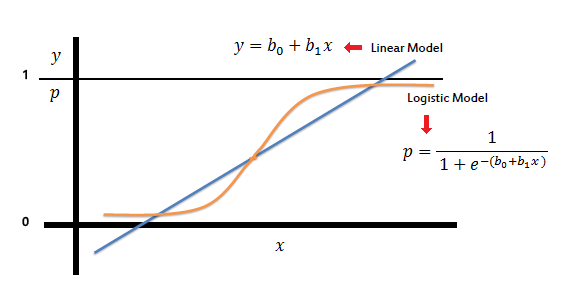}
            \caption{Data Mining graph of Logistic Regression}
            \label{fig:Logistic_Regression}
\end{figure}

\end{enumerate}

\section{Feature Level and Decision Level Fusion}
Multimodal fusion is the heart of any multimodal sentiment analysis engine. There are two main fusion techniques: feature-level fusion and decision-level fusion. Feature-level fusion is implemented by concatenating the feature vectors of all three modalities to form a single long feature vector. Despite its simplicity, this method produces accurate results. We concatenated the feature vector of each modality into a single feature vector stream. This feature vector is used for classifying each video segment into sentiment classes. To estimate the accuracy, we used tenfold cross-validation. 

In decision-level fusion, we obtained feature vectors instead of concatenating the feature vectors, as in feature-level fusion, where we used a separate classifier for each modality. The output of each classifier is treated as a classification score. We obtained a probability score for each sentiment class from each classifier. As there are three sentiment classes, we obtained three probability scores from each modality.   

\section{Experimental Setup}
The objectives of Human-Computer Interaction are to create usable and safe systems, as well as functional systems. To create computer frameworks with great ease of use, developers must endeavor to comprehend the factors. 

\begin{enumerate}
\item that decides how people utilize technology, 
\item create tools and procedures to empower existing systems, 
\item accomplish productive, powerful, and safe collaboration 
\item put users first 
\end{enumerate}

\begin{table}[h!]
\centering
\caption{Mapping of Gestures with Arduino Boards}
\label{table:1}
\begin{tabular}{|c  c|} 
 \hline
  Fist  & Pin 3\\
 \hline
 Wave In & Pin 4\\
\hline
Wave Out & Pin 5 \\
\hline
Finger Spread & Pin 9\\
\hline
Double Tap & Pin 10 \\
\hline
\end{tabular}
\end{table}

Underlying the whole theme of HCI is the conviction that how people utilize a computer system. Their necessities, abilities, and inclinations for leading different assignments should direct the developers in how they outline the systems. People should not have to change how they use a system to fit in with it. Rather, the system should be designed to coordinate their requirements \cite{Model:online}. The continuous advancement in technology requires machines meant for different purposes, and the recent trend involves human participation \cite{urban2005fusion}. Human-robot communication is conceivable through two strategies.

\begin{itemize}
\item Accepting user input from peripheral gadgets that are autonomous of each other, and 
\item Accepting user input through various modalities and intertwining them as a method for acquiring the semantics related to the activities of the user.
\end{itemize}

The second approach is the focal point of this paper. In 2005, a framework was composed which acknowledges contribution in the form of speech, keystrokes, and motions. This framework could resolve questionable sources of info and organize them \cite{turk2014multimodal}. The combination of multiple inputs utilized in various applications of daily use and its extension isn't just kept to robots, for example, combining speech and facial identification. A framework was composed in 1999 that could verify a client by looking at contributions against a pre-populated database \cite{thomas1989problem}. The most recent change in Microsoft Windows is that it is capable of validating clients through a webcam connected to the computer system \cite{tanaka2002multimodal}, even though it is an unimodal framework that could be upgraded with more modalities to enhance its precision and make it less vulnerable to outside attacks or spoofing. The utilization of EMG data for fusion is rarely encountered in any real-life application; one such application was executed to control electronic musical devices through EMG and relative position sensing \cite{schwab2017fourth}. The concept of multimodal data fusion has been executed in mechanical robots utilizing the Microsoft Kinect and sensor equipment called Asus Xtion Monitor by catching hand gestures identified by two Leap Motion sensors and playing out the subsequent mapped activities on a robotic arm \cite{salman2014multi}. During the most recent five years, human-robot interaction has mainly been performed utilizing the Microsoft Kinect system; not many multimodal framework plans have utilized EMG data to intertwine with speech, text, and other different modalities.

\begin{figure}[h!]
            \centering
            \includegraphics[width=0.4\textwidth]{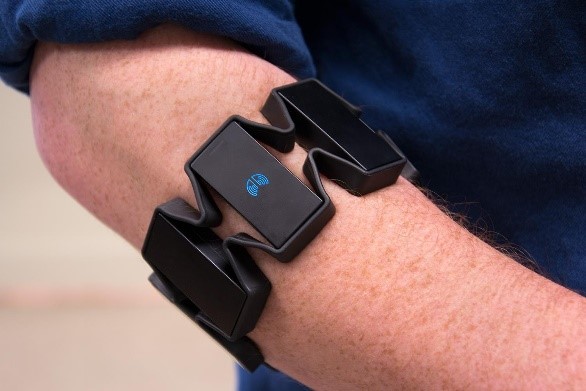}
            \caption{MYO Arm Band}
            \label{fig:MYO}
\end{figure}

The Microsft Kinect is equipped for capturing both voice and signals on an independent basis. Also, there is no feature to capture the EMG data of human limbs using Kinect. This constraint drove us to another special innovation called the MYO sensor armband. We have chosen to utilize it as one of the modalities and perform fusion to enhance the accuracy and performance. MYO armband is open-source programming that gives roads to tweak gestures and utilize them in gadgets utilized as a part of everyday life, for instance, controlling a wheelchair, turning a doorknob, and so forth \cite{gabriel2016gesture}. The outcomes accomplished with different experiments to assess the precision of MYO turn out to be 87.8 to 89.38\%, which gives us avenues of improvement \cite{luh2016muscle}. MYO band is utilized as a part of an experiment to perform both search and select operations on a PC, and the average score for assessment is analyzed; the analysts infer that after embracing the limb's temperature, the MYO band continually performs with a similar number of scores \cite{csapo2016evaluation}. To the best of our knowledge, this is the first research work that utilizes EMG data to catch gestures utilizing MYO armband sensors for Multimodal data fusion.

\begin{table}[h!]
\centering
\caption{Preliminary results for of Muscle sensor MYO band}
\label{table:2}
\begin{tabular}{|c  c c |} 
 \hline
 \textbf{Gesture} & \textbf{Wrong/Missed gesture \%} & \textbf{Correct \%}\\
 \hline
  Wave Out  & 9.5 & 90.5\\
 \hline
Wave In & 	9.1	& 90.9\\
\hline
Fist & 	13.6	& 86.4\\
\hline
Double Tap	& 20.6	& 79.4\\
\hline
Finger Spread	& 14.5 &	85.5\\

\hline
\end{tabular}
\end{table}

\begin{table}[h!]
\centering
\caption{Preliminary results for Microsoft Speech used by non-native speakers.}
\label{table:3}
\begin{tabular}{|c  c  c  |} 
 \hline
 \textbf{Command} &	\textbf{Microsoft Speech API (Wrong output)	}& \textbf{Correct}\\
 \hline
Move Right & 	10	 & 90\\
 \hline
Move Left &	34.2	 & 65.8\\
\hline
Move Up & 	8.9	& 91.1 \\
\hline
Move Down & 	22.5	& 77.5 \\
\hline
Move Gripper &	14.1 &	85.9\\
\hline
\end{tabular}
\end{table}

\section{ Methodology}
The problems which have been tried to be solved deal with decision-making techniques when the system receives ambiguous user input. There are several machine learning algorithms available that help the model to learn the behavior of the user. 

The robots initially introduced to the market were generally basic, the majority of them requiring a teaching phase and programming. Off lately, the robots have become dynamic, modern, and significantly more proficient than before \cite{roitberg2015multimodal}. Along with this modern sophistication came expanding requests to perform complex tasks that require both precision and accuracy. The robot used for the experiment in this paper opens the adequate opportunity to get better in both robustness and accuracy. Henceforth, it was chosen to enhance the precision of a robotic arm by the utilization of multi-modal data fusion. The experiment emphasizes the change of information provided through various channels into a single format which is comprehended by the robotic arm through mediation. The input modalities utilized are speech and gesture.

\subsection{Hardware}

\begin{figure}[h!]
            \centering
            \includegraphics[width=0.4\textwidth]{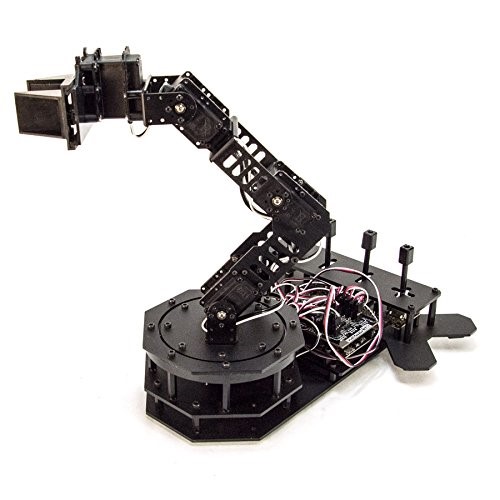}
            \caption{Arduino based robotic arm.}
            \label{fig:Arduino}
\end{figure}

\begin{figure}[h!]
            \centering
            \includegraphics[width=0.4\textwidth]{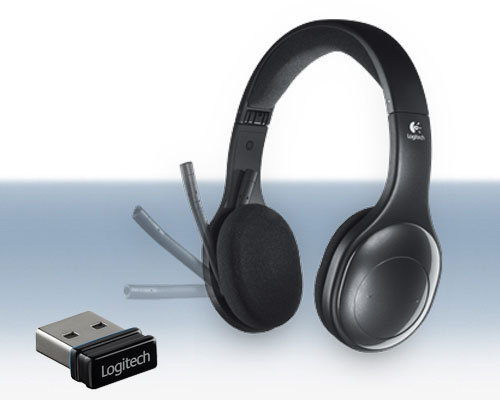}
            \caption{Logitech Wireless Headset H800}
            \label{fig:Logitech}
\end{figure}

The framework designed for the experiment depicted in this paper is made out of the accompanying segments: An Arduino-based robotic arm and a MYO armband. These gadgets are outlined in \ref{fig:MYO} and \ref{fig:Arduino} separately. The robotic arm utilized is made by Trossen Robotics. The robot utilized as a part of the analysis portrayed here is called 'RobotGeek Snapper Arduino Robotic Arm,' and it contains five servo motors. An electromyography data-based sensor controls the robotic arm called a MYO. \ref{fig:MYO} portrays the utilization of the MYO band from which information is captured and manipulated to perform activities on different Arduino-based devices. The armband is fit for catching five signals: Fist, Wave Left, Wave Right, Double Tap, and Fingers Spread, as appeared in \ref{fig:MYO_Gestures}. This armband allows tweaking an open library and perform activities per requirement. The robotic arm utilized involves Arduino Duemilanove and Diecimila microcontroller board for accepting inputs through USB. The Arduino board associated with the robot with pins is characterized for every servo motor, as shown in \ref{fig:Arduino}. A high-precision wireless H800 headset from Logitech \ref{fig:Logitech} was utilized for capturing the human speech input.

\subsection{Software}

The software is implemented utilizing C\# and
 C++ programming languages. C++ is utilized to execute the MYO API, while C\# is utilized for giving speech input utilizing Microsoft's speech API. At first, the robotic arm was operated using gestures only. The MYO armband was utilized to calculate the robot's accuracy. It is presumed that the accuracy of the armband was not high. To enhance the precision of the commands, an imparted fusion of the data is incorporated. The speech modality is clubbed with the gestures defined in the MYO Armband API. The MYO armband gives an API to control the Arduino board, which is connected to the robotic arm, as appeared in Figure 5. The Arduino board comprises 14 pins and can interface with every servo motor controlling the robot.

\subsection{Experiments performed}
\begin{figure}[h!]
            \centering
            \includegraphics[width=0.4\textwidth]{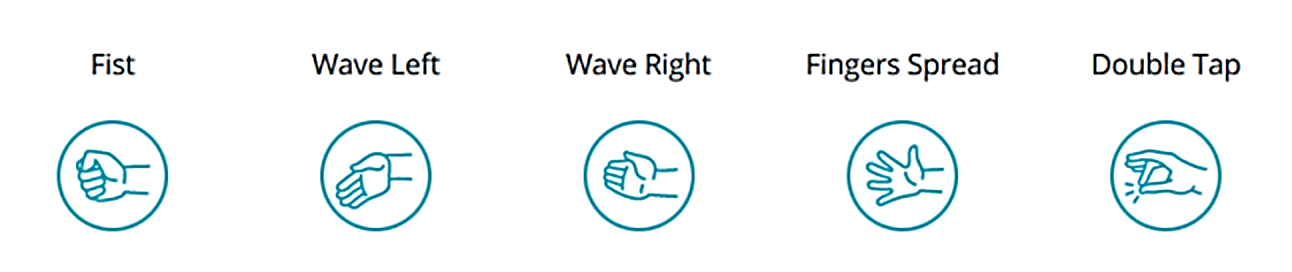}
            \caption{MYO Gestures}
            \label{fig:MYO_Gestures}
\end{figure}

The experimental framework is planned to utilize a client-server paradigm. The MYO armband comprises eight sensors that capture muscle development. Such sensors contrast and match hand developments with motions characterized in MYO. For instance, the Wave Out motion recorded in \ref{table:2} was performed by moving the hand in the vertical direction in the third gesture of \ref{fig:MYO_Gestures}. The MYO program is working as the server while the Microsoft speech program is actualized as the client.

As mentioned above, the precision of capturing gestures isn't high, which tends to bring about matching the hand development to the wrong motion; e.g., Wave Out might be recorded as Wave In. Additionally, the band, once in a while neglects to record the gestures completely. Both cases are considered errors. The MYO API permits customization as per the user's requirement, and the prototype is implemented using threads responsible for listening to gestures. If the armband misses a gesture, voice input compensates for the missed information through human speech. Handling of speech input implemented in a client part, which sends commands to the server (MYO API). Priority is assigned to the MYO band; however, on account of an error, speech recognition actuates and helps enhance the accuracy of controlling the robot. Fusion is thus performed in the order of priority. Gestures were given the highest priority. If an occurrence of an inability to catch the input should arise, voice commands are utilized to redress and fill in as the main input. Priority-based fusion is utilized in different domains, including medical systems, and tends to enhance its precision significantly \cite{ben1999fusion}.

\begin{figure}[h!]
            \centering
            \includegraphics[width=0.4\textwidth]{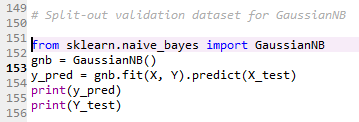}
            \caption{Split Out Validation Dataset for Gaussian NB}
            \label{fig:Split_Out}
\end{figure}

The speech fused with EMG input received from MYO, which enables the robot to work precisely as per the user’s input command. When the user performs gestures using his/her arm, the input message is transmitted from the MYO band to the Arduino, and as a result, it moves the specific servo motor. The fused input sets the corresponding Arduino pin to high, i.e., 1, which then moves the robot. In the prototype constructed, five different Arduino pins were linked to various gestures as shown in \ref{tableThank :1}. Through gestures and speech, users should be able to control the robotic arm precisely and accurately; the results are shown in the next section.

We have implemented the machine learning algorithm with our test data and evaluated the results. The result explains that the K-neighbor classifier yields the most accurate results among all the algorithms. In our earlier results, the results were based on input from different modalities.

\section{Results and Discussion}

A performance evaluation was executed, keeping in mind the end goal to evaluate how exactly the modalities are independent; subsequently, we tried them using machine learning techniques. The complete details of the results are outlined in Table \ref{table:13}. The test results demonstrate that the algorithm that works best for us is the K-Neighbors Classifier, with an accuracy of 92.45 percent.

\begin{table}[h!]
\centering
\caption{Precision, Recall and F1-Score}
\label{table:4}
\begin{tabular}{|c c c c c|} 
 \hline

 \textbf{ML Algorithm} & \textbf{Precision} & \textbf{Recall} & \textbf{F1-Score} & \textbf{Support} \\
\hline
SVM & 0.82 & 1.00 & 0.90 & 9 \\
\hline
Gaussian NB & 0.67 & 0.50 & 0.57 & 4 \\
\hline
Decision Tree Classifier & 0.83 & 0.83 & 0.83 &	6 \\
\hline
Lin. Discriminant Analysis & 1.00 & 0.75 & 0.86 & 8 \\
\hline
Logistic Regression & 0.78 & 1.00 & 0.88 & 7 \\
\hline
KNN & 0.92 & 0.92 &	0.91 & 1 \\
\hline
\end{tabular}
\end{table}

Similarly, the MYO band results were captured to quantify accuracy and to find the scope of improvement \cite{mohd2017multi}. The preliminary results have shown the MYO band has a scope of improvement. The error rate lies between 9.1 and 20.6 percent. The data has been collected by experimenting ten times, with each experiment having one hundred gestures performed and then calculating the average percentages shown in \ref{table:10}. An error in the experiment occurs when a gesture is either missed or captured wrong. The trials have been performed in laboratory conditions. The MYO armband is capable of adapting to specific human limbs and improves its output once it has been trained completely.

\begin{table}[h!]
\centering
\caption{Fusion results with Error \% \& Variance}
\label{table:5}
\begin{tabular}{|c  c c c c c c  |} 
 \hline
 
 \textbf{Fusion} & \textbf{50} & \textbf{100} & \textbf{150} & \textbf{200} & \textbf{Err}& \textbf{Var} \\
 \hline
  Move Gripper \& Double Tap  & 7	& 2 & 	2 & 	4	& 7.5 & 	5.58  \\
 \hline
Move Down \& Fist & 	3	& 1 & 	2	& 2	& 4	& 0.67\\
\hline
Move Up \& Finger spread &	3 & 	3 & 	2 &	2 &	5 & 	0.33  \\
\hline
Move Left \& Wave left	& 0 & 	3 & 	2	& 2 & 	3.5 & 	1.58 \\
\hline
Move Right \& Wave out &	3 &	3 &	4 &	2 & 	6 & 	0.67\\

\hline
\end{tabular}
\end{table}

The armband sensors turn warm soon after it is worn on the arm of a user, and thus adapting to body temperature and then capture the gestures precisely after 1-2 minutes. If one's arm is cold, the sensors are unable to capture the gestures precisely.
The Microsoft API results are assessed by speaking a command, and if the command is captured incorrectly, the instance is marked as an error. Incorrect capture is defined as the resulting string being caught twice or having incidental words or characters added to it. 

\begin{figure}[h!]
            \centering
            \includegraphics[width=0.4\textwidth]{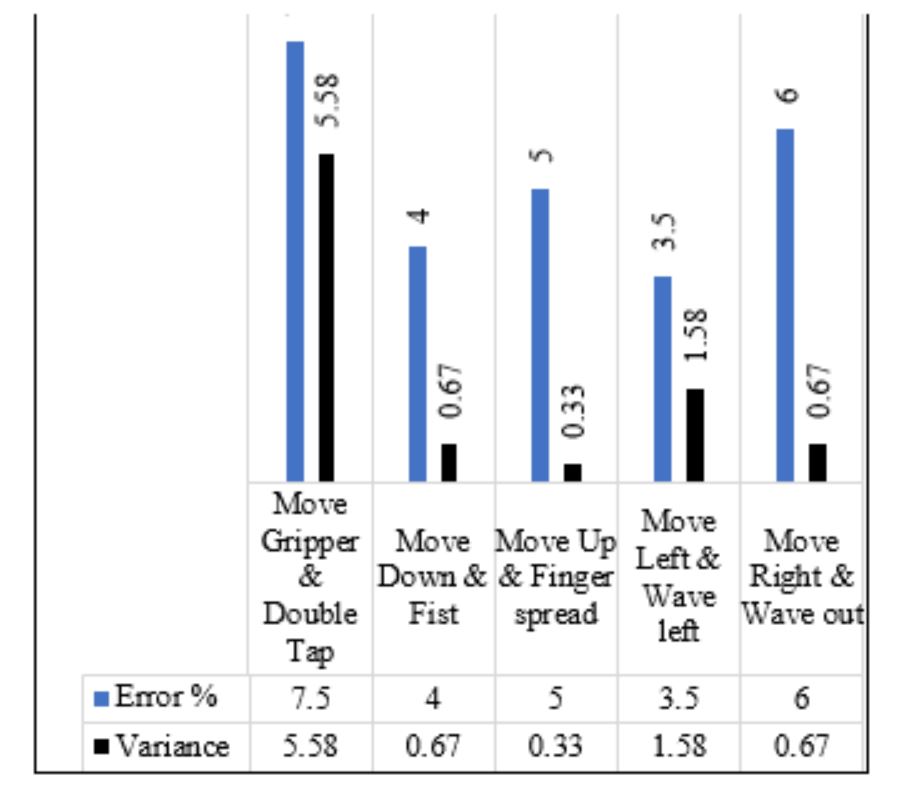}
            \caption{Error \% deviation of fused Inputs}
            \label{fig:Error_Deviation}
\end{figure}

Multimodal data fusion of voice and motion utilizing the MYO band enhances the system performance significantly. The test results are shown in \ref{table:4}. After the implementation of the fusion of the robotic arm, the error rate decreased to 5.2\%, which is an average of all errors. The fluctuation of the error rate is shown in \ref{fig:Error_Deviation}. The errors are, for the most part, because of reading the wrong gesture, e.g., the finger spread sometimes captured as a fist, which is considered as an error. Experiments are performed on all five fusion input tests 200 times each, and the percentages are calculated respectively.

\section{Limitations}
Speech and electromyography data were the modalities used in the system constructed. The MYO band used for capturing the EMG data is capable of recognizing five gestures. This limited the number of operations that could be performed on the robotic arm. The second challenge lies in capturing the speech commands using the Microsoft Speech API. Non-native speakers of the English language will face difficulties and challenges in approximating their accent to that of a native speaker. This created difficulty in conveying commands correctly. Four areas need to be worked on and improved regarding human-robot interaction. These include speech localization, language understanding, dialogue management, and speech synthesis \cite{judd1990technique}. Also, as the ultimate goal of this research is to improve accuracy, the approach here described map commands to all possible options that the Microsoft Speech API recognizes as valid (for example, “move right” sometimes gets recognized as “override” – an incorrect response). This provided us with a way to quantify the accuracy of the system. We prepared a many-to-one mapping of all these possible combinations to a particular voice command. 

\section{Conclusion and Future Work}
The implementation of machine learning algorithms provides us with results with an accuracy of 90.56 percent. The future idea is to design a model using a Google Speech API, which processes the complete natural language and extracts meaning from it. When the user provides the statement ‘move the robotic arm ten degrees to the left,’ the system should capture the command and process it with the keywords ‘left’ and ‘ten degrees.’ The idea which shall be implemented is to process the string and pass the parameters to the server program. The server program captures the parameters and processes the command on the robotic arm. In the future, we would like to extend our work while incorporating the complete natural language commands using Google API, which allows the users to communicate and customize the input as per their needs. The designed system should be capable enough to learn commands using machine learning techniques.

\bibliographystyle{ieeetr}
\bibliography{references}

\end{document}